\documentclass[review,12pt,authoryear]{elsarticle}
\usepackage{framed,multirow}
\usepackage{fullpage}

\usepackage{nomencl} 
\makenomenclature
\renewcommand*\nompreamble{\begin{multicols}{2}}
\renewcommand*\nompostamble{\end{multicols}}
\usepackage{amssymb}
\usepackage{amsmath}
\usepackage{graphicx}
\usepackage[retainorgcmds]{IEEEtrantools}
\usepackage[colorlinks]{hyperref}
\hypersetup{
  citecolor = {blue}
}
\usepackage{siunitx}
\usepackage{pdflscape}

\usepackage{enumitem}
\usepackage[normalem]{ulem}
\useunder{\uline}{\ul}{}
\usepackage{setspace} 
\usepackage[utf8]{inputenc}
\usepackage[linesnumbered, ruled, vlined]{algorithm2e}
\usepackage{rotating}
\usepackage{nicefrac}
\usepackage{xspace}
\usepackage{float}
\usepackage{caption}
\usepackage{subcaption}
\usepackage{rotating, lscape, longtable, tabu, booktabs, siunitx, multirow}
\usepackage[capitalise, noabbrev]{cleveref}
\usepackage{verbatim}
\usepackage{mathtools}
\usepackage{makecell}
\usepackage{lineno}
\usepackage{threeparttable}
\makeatletter
\def\ps@pprintTitle{%
 \let\@oddhead\@empty
 \let\@evenhead\@empty
 \def\@oddfoot{}%
 \let\@evenfoot\@oddfoot}
\makeatother
\begin{document}

\begin{frontmatter}

 \title{Evaluating sampling strategies for effective detection of African swine fever in growing pig population in the U.S.}%
 \author[cvm]{Jason A Galvis\corref{cor2}}
 \author[cvm]{Aniruddha Deka\corref{cor2}}
 \author[cvm]{Gustavo Machado\corref{cor1}}
 \ead{gmachad@ncsu.edu}

 \address[cvm]{Department of Population Health and Pathobiology, North Carolina State University, Raleigh, North Carolina, USA.}

 \cortext[cor2]{These authors contributed equally to this work.}
 \cortext[cor1]{Corresponding author:}

\begin{abstract}

Early detection of African swine fever virus (ASFV) is critical to preventing widespread epidemics. However, the effectiveness of within-farm sampling remains to be examined, particularly during the early phase of an outbreak when disease prevalence is low, animals may be asymptomatic, or clinical signs are masked by co-circulating diseases. This study assessed four sampling strategies for detecting ASFV-infected animals in suspected growing pig farms within the first 14 days of the introduction of either a high- or moderate-virulence ASFV strain. Pens were selected using three methods: random sampling, targeted sampling of pens with clinical animals, and informative sampling based on estimated pen infection probabilities. The informative sampling method was further divided into sequential method, which ranked pens by descending viral load probability, and cluster \& random method, which selected pens at random from high and low viral load clusters. Each pen-selection method was examined with different sample sizes. We calculated the sensitivity of each approach as the probability of detecting at least one ASFV-positive pig per farm. Results showed that sampling 30 pens with one pig per pen using the target \& random pen-selection method yielded the highest detection sensitivity, even in the presence of other co-circulating diseases that interfere with the accurate identification of clinical ASFV cases. In contrast, sampling five pens resulted in the lowest sensitivity. These findings provide valuable insights for improving ASFV surveillance strategies in the U.S. and can inform preparedness efforts for other foreign animal diseases.

\end{abstract}




\end{frontmatter}
\section{Introduction}
African swine fever virus (ASFV) is an infectious pathogen that affects domestic swine and wild pigs \citep{gallardo_african_2015, sanchez-vizcaino_update_2015}. ASFV strains vary in virulence, categorized as high, moderate, and low, which significantly influence the virus' transmission, detection dynamic and the severity of the disease, each strain presenting unique challenges for control efforts \citep{galindo_african_2017, sanchez-cordon_african_2018, dixon_african_2019}. While ASFV has not been detected in the U.S., its future presence in the North America raises significant concerns \citep{schambow_update_2025}. If introduced into the U.S., the virus could severely impact the national swine industry, highlighting the urgent need for robust surveillance and early detection strategies to mitigate potential economic losses and safeguard business continuity \citep{sykes_identifying_2025, galvis2025estimating}.

In North America commercial growing pig farms, pigs are grouped in pens, with pens comprising a room and one or more rooms comprising a barn. The current U.S. ASFV surveillance strategies for suspected farms involve sampling a subset of pens and pigs throughout the farm \citep{galvis2025estimating, usda_african_2023}. While there is no clear guidance in the national response plan \citep{usda_african_2023} on the exact number of pens and animals to sample, by discussions with the swine industry and state animal health officials it has been established that five pigs per pen as feasible number \citep{galvis2025estimating}, prioritizing pens with pigs exhibiting clinical signs \citep{usda_african_2023}. However, farms in the early stage of infection with the absence of clinical animals, pens selection would be random. Random pen sampling may miss early-stage infections \citep{murato_evaluation_2020,robert2024oral}, 
while targeted pen sampling based solely on clinical signs does not account for subclinical infections \citep{gallardo_african_2019, trevisan_active_2024, bonney_simulation_2024}. Any delay in detecting early infection can lead to the further spread of the virus within and between farms, impacting control efforts and exacerbating economic losses \citep{sykes_estimating_2023}. Thus, there is a critical need to evaluate and refine sampling pen methods to enhance early detection \citep{murato_evaluation_2020}.

Another key challenge in clinical sign-based sampling is the presence of other diseases that may confound the subjective assessments of which pig, pen or barn should be sampled. This issue is especially relevant in the early stages of an ASFV outbreak when the number of infected animals and mortality within expected ranges and non-specific lesions are common, making it difficult to differentiate ASFV from other swine diseases \citep{sanchez-vizcaino_update_2015,nga2020clinical,faverjon2021risk}. In the U.S., endemic pathogens such as porcine epidemic diarrhea virus (PEDV) and porcine reproductive and respiratory syndrome virus (PRRSV) are widespread among breeding and growing pig farms \citep{niederwerder_swine_2018, neumann_assessment_2005}. Although these diseases have distinct clinical characteristics, differential diagnostics from ASFV are necessary, in particular for acute PRRSV outbreaks. For example, pigs infected with PRRSV can exhibit lethargy, cyanosis (a.k.a. blue skin), respiratory distress, abortion, and loss of appetite, symptoms similar to those of ASFV \citep{sanchez-vizcaino_update_2015, yoon_clinical_2020,penrith2024african,nishi2022establishment}. As a result, PRRSV-infected pigs may be mistakenly selected for ASFV sampling, reducing the likelihood of detecting ASFV \citep{guinat_english_2016}. Therefore, it is essential to assess strategies that enhance the probability of correctly identifying ASFV-infected pigs in the presence of co-circulating diseases.

ASFV surveillance in North America lacks a clearly defined number of pens or animals that should be sampled when a farm is suspected to be positive \citep{murato_evaluation_2020}. The uncertainty about the optimal sampling schema risks the effectiveness of early detection and containment efforts, as an inadequate sampling size may fail to identify infected animals \citep{atuhaire_prevalence_2013, bonney_simulation_2024}, allowing the disease to spread unnoticed. Establishing a standardized approach to determine the optimal number of pens and animals to sample is critical for enhancing future surveillance efforts' efficiency. Previous studies have simulated and evaluated the performance of sampling strategies to detect ASFV-positive farms \citep{malladi2022predicting, halasa_simulating_2016}. Recently, Bonney et. al. 2024 evaluated premovement sampling strategies for finisher farms by simulating disease dynamics at the animal level across three different barn sizes, assessing detection strategies based on time intervals, and targeting either clinical pigs or random pigs in the absence of clinical cases \citep{bonney_simulation_2024}. While this study provided valuable information for ASFV detection, there remains a question about more precise sample strategies that could identify pens with a higher probability of disease infection in the absence of clinical animals. Furthermore, a significant limitation also remains in most ASFV simulation studies, in which estimations were based on simulated or a small subset of the available data \citep{bonney_simulation_2024,hayes_mechanistic_2021}. To provide comprehensive results, estimations need to be performed at large scale with data that represent the diverse U.S. swine industry. This would require comprehensive farm demographic data, including barn sizes and the number of pens, as these factors are critical to designing effective sampling strategies at national-level \citep{deka2025modeling, safari_modeling_2024}.

Ultimately, effective on-farm ASFV surveillance should balance detection sensitivity and resources to sample an adequate number of animals to prevent false negatives delaying early detection and lead to large outbreaks \citep{sykes_estimating_2023}. To achieve this balance, it is essential to exanimate the effectiveness of various sampling methods under different epidemiological scenarios \citep{murato_evaluation_2020, galvis2025estimating}. 
This study systematically evaluates multiple sampling methods for detecting ASFV-positive animals in suspected farms through a simulated ASFV outbreak. It considers high- and moderate-ASF virulence strains, PRRSV as a co-circulating disease, four pen-sampling approaches, and variations in pig and pen sample sizes across 1,865 commercial swine farms in 33 U.S. states. The model outputs determine the optimal number of pigs and pens to sample, maximizing sensitivity for detecting true ASFV-positive farms. These findings provide valuable insights to optimize ASFV surveillance protocols and support the swine industry's business continuity.

\section{Methodology}

\subsection{Data}
This study used data from 1,865 growing pig farms managed by 48 swine production companies across 33 U.S. states. The different barn layouts were accessed using farm maps from the enhanced on-farm Secure Pork Supply (SPS) biosecurity plans, from the Rapid Access Biosecurity application (RABapp™) \citep{machado_rapid_2023}. The \href{https://www.securepork.org/}{SPS biosecurity plans} are part of a USDA and Pork Checkoff initiative to improve business continuity by helping swine producers implement robust biosecurity measures. Each plan includes a site-specific biosecurity strategy and a detailed farm map featuring fourteen elements, including the line of separation (LOS) \citep{fleming2025enhancingusswinefarm}. The LOS is a control boundary to prevent pathogen movement into barns where susceptible animals are housed. In RABapp™, these farm maps are stored as geospatial data, with LOS polygons used to define each barn's boundaries. This enabled us to calculate barn sizes and approximate the number of rooms and pens within each barn (Supplementary Material Section 1). From the 1,865 farms, we estimated 7,317 barns with a total of 7,704 rooms, where  90\% of the barns had one room and 10\% two rooms. Given the standardized pen sizes across the industry, where the average pen dimensions are 19.6 ft x 9.8 ft, housing 15-35 pigs per pen \citep{muirhead2013managing}, we developed an algorithm to generate pen distributions within each room based on the pen area \citep{deka2025modeling}. The number of pigs per farm documented in the SPS plans and stored in RABapp™ was used to allocate pigs to pens by distributing the total number of pigs across the total estimated number of pens on each farm.

\subsection{ASFV transmission model}

\begin{figure*}[!htb]
    \includegraphics[width=\linewidth]{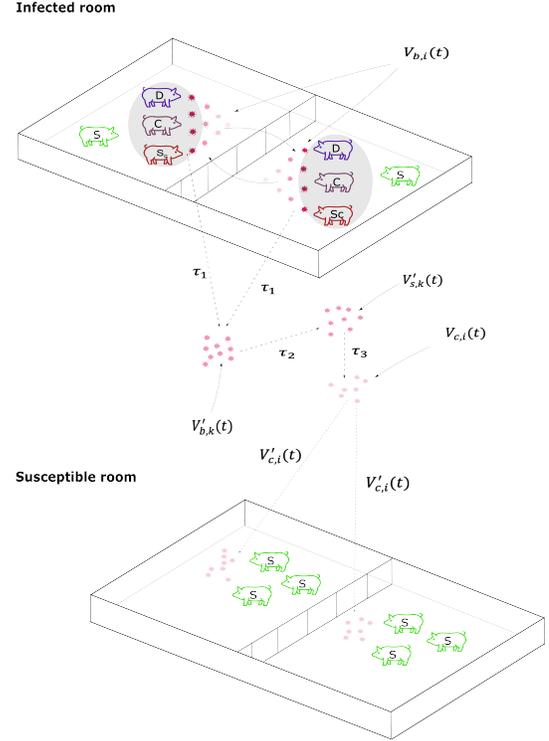}
    \caption{\textbf{African swine fever transmission model framework}. (A) Schematic illustrating the transition of pigs to different compartments: Susceptible ($S$), Exposed ($E$), Clinical ($C$), Sub-Clinical ($S_c$), Carrier ($C_c$), Detected ($D$). $\lambda_i$ represents the force of infection, indicating the rate at which pigs transition from the Susceptible to the Exposed compartment. $i=1,2,...,7$ defines the various pathways through which exposure can occur. $\theta=1,2,3$, specifies the transitions of Exposed pigs to Clinical, Sub-Clinical, or Chronic-Carrier compartments after the latency period. $\kappa$ describes the status changes of pigs within the Clinical, Sub-Clinical, and Carrier compartments. $\delta$ is the detection rate of Infected pigs, encompassing those in the Clinical, Sub-Clinical, and Carrier compartments. (B) We illustrate the seven transmission routes: Direct nose-to-nose contact within pens, within-pen fecal transmission rate, within-pen aerosol transmission, between-pen direct nose-to-nose contact, between-pen fecal transmission rate, between-pen airborne transmission, and human transmission. (C) Between-room transmission by environmental contamination around each farm.
    Here, $V_{b,i}(t)$ represents the cumulative viral load accumulated inside a room from all the pens at time $t$. The parameter $\tau_1$ denotes the fraction of particles that disperse from the room. $V'_{b,k}(t)$ represents the cumulative particles in the environment accumulated from all the rooms. The parameter $\tau_2$ is the survival rate of particles in the environment, and $V_{c,i}(t)$ represents the particles that have survived in the environment. Furthermore, $\tau_3$ denotes the fraction of particles that enter from the environment, while $V'_{c,i}(t)$ represents the particles that are distributed to each pen.
    } \label{fig:status_and_routes}
\end{figure*}

We developed a stochastic, animal-level transmission model to simulate ASFV dynamics for high and moderate virulence strains within growing pig farms \citep{deka2025modeling}. Our model divided the swine population into six distinct health states for ASFV: Susceptible $(S)$, Exposed $(E)$, Clinical $(C)$, Sub-Clinical $(S_c)$, Chronic Carrier $(C_c)$, and Detected $(D)$ (Figure \ref{fig:status_and_routes}A). Additionally, we incorporated three states to represent the co-circulation of PRRSV: Susceptible (ASFV)-PRRSV (Infected) $(S_p)$, Exposed (ASFV)-
PRRSV (Infected) $(E_p)$ and Clinical (ASFV)-PRRSV (Infected) $(C_p)$  (Figure \ref{fig:status_and_routes}A). Pigs in each state had a probability of dying $\mu_i , \text{where } i = s, s_p,s_e,s_c e, s_c, c, c_c,$ and  $d$, at different rates.

Typically, PRRSV-infected pigs remain viremic for three or four weeks \citep{zimmerman_porcine_2019}. Considering this timeframe, we assumed that pigs infected with PRRSV remain clinical throughout the simulated period (14 days), establishing a fixed PRRSV prevalence within rooms and limiting the transmission of PRRSV between animals. Based on this premise and assuming a hypothetical prevalence, 10\% of the total pigs in each room were randomly seeded as PRRSV clinical ($S_p$) and distributed across the pens at the beginning of the simulation. Pigs in the $S$ state transited to $E$ through direct and indirect transmission routes, categorized into four levels: within-pen, between-pen, within-room, and between-room levels. The within-pen transmission included nose-to-nose contact, airborne transmission, and oro-fecal transmission between the pigs within the same pens \citep{guinat_transmission_2016,deka2025modeling}. Between-pen transmission encompassed transmission between adjacent pens via direct contact, fecal contamination, and room-level airborne transmission \citep{olesen_transmission_2017, safari_modeling_2024,deka2025modeling}. Within-room transmission incorporated human-mediated transmission, where barn personnel inadvertently spread the virus between pens through contaminated clothing, equipment, and movement \citep{chenais_epidemiological_2019, deka2025modeling}. Between-room transmission included airborne pathogens released into the environment, which is allowed to contaminate all rooms present on the farm. Additionally, the model accounted for the movement of pigs between pens, simulating standard swine management practices in which pigs are sorted and grouped together based on weight and health \citep{fraser_general_2013}. The total force of infection $\lambda$ was calculated by summing the force of infection from each route, determining the rate at which Susceptible pigs become Exposed. Section 2 of the Supplementary Material provides a summary of all transmission routes within a room.

For the between-room transmission, each room housing infected pigs accumulated viral load at each time step, denoted as $(V'_{b,k}(t))$, $k=1,2,3,..$ was the number of rooms of each farm, with a fraction ($\tau_1$) of these particles dispersing into the environment. The total environmental viral load outside a farm $(V_{\text{pixel}}(t))$ was obtained by aggregating viral particles across all rooms, where particles persisted at a survival rate ($\tau_2$), determining the total environmental particle load $(V'_{s,k}(t))$. A fraction of these particles was then reintroduced into barns and evenly distributed across all rooms as $(V_{c,i}(t))$. Within each room, the reintroduced particles follow airflow-driven movement, captured by the normalized transmission matrix \(\tilde{G}_i\). The number of viral particles reaching pen \(i\) was then determined by $V'_{c,i}(t)$. Thus, the likelihood of a susceptible pig becoming exposed via airborne transmission was governed by the combined effect of room-level viral load and environmental reintroduction. Additional details about the environmental transmission between rooms can be found in Supplementary Material Section 2. Exposed pigs can transit to Clinical, Sub-Clinical, or Chronic Carriers based on a latency period rate $\theta_1$, $\theta_2$, and $\theta_3$, respectively (Supplementary Material Table S1 and S2).

\subsection{Surveillance assumptions and outbreak scenario}

In this study, we assumed an active ASFV outbreak in the U.S., and stakeholders were aware of it. Thus, farms could be detected either through farm level efforts (internal farm surveillance) or through coordinated sampling by official authorities (suspected farm surveillance). For suspected farm surveillance, we assumed that all farms were suspected because they were either located within an ASFV control zone or had contact with a detected farm, thereby triggering surveillance activities \citep{usda_african_2023}. To evaluate the effectiveness of each surveillance strategy independently, we assumed that farms detected through internal farm surveillance remained undetected for the purposes of assessing detection through suspected farm surveillance. We assumed that farms became infected at the same time they were identified as suspected. Finally, although the sampling of dead pigs is one of the most effective detection strategies \citep{usda_african_2023}, this study focused on sampling live pigs to simulate the logistical and diagnostic challenges associated with this approach.

\subsection{Internal farm surveillance: Regular detection by stakeholders}

Clinical, Sub-Clinical, or Chronic Carrier pigs had a daily probability $\nu$ to be detected through passive surveillance, defined here as the identification of infected pigs via routine observation of clinical signs or incidental findings during non-targeted diagnostic activities (e.g., detection during testing originally for another disease, such as PRRSV). For the highly virulent ASFV strain, we assumed that $\nu$ follows a PERT distribution with a mean of 0.0045 (range: 0.001–0.009). For the moderate-virulence strain following a subacute disease course, we assumed a distribution with values approximately half of those for the high-virulence strain, with a mean of 0.0023 (range: 0.0005–0.0045), reflecting milder and more difficult-to-recognize clinical signs through routine observation \citep{gallardo_african_2015, sanchez-vizcaino_update_2015}.

We simulated switching from passive to active surveillance through two mechanisms: i) pig mortality and ii) suspected farm being reported (described in detail section \ref{subsec:suspected_farm_surveillance}). Pig mortality triggers active internal farm surveillance if the number of dead pigs exceeded 2\% of the room population within a 14-day period (approximately two to four times the typical mortality observed in U.S. finisher farms within two weeks) \citep{losinger_mortality_1999, stalder_pork_2013}. We defined this mortality trigger surveillance as targeted detection efforts initiated in response to abnormal mortality patterns, leading to intensive clinical inspections and diagnostic testing \citep{barongo2016mathematical,dankwa2022stochastic}. Here, the daily probability of disease detection increases to 0.95 for Clinical pigs, 0.75 for Sub-Clinical, and 0.5 for Chronic Carriers for high and moderate virulent strains (Supplementary Material Table S1 \& Table S2). Once a pig was detected, the entire farm was considered detected.

\subsection{Suspected farm surveillance: Pig sampling and testing by the animal health official service}
\label{subsec:suspected_farm_surveillance}
Active surveillance by suspected farm reports was triggered when farms were in contact with an ASFV-positive farm by animal or vehicle movements or when located within an infected, buffer or surveillance zone around an ASFV-positive farm \citep{sykes_estimating_2023, usda_african_2020}. This strategy included collecting blood samples from a set of pigs within the farm \citep{usda_african_2023}. For the sampled pigs, we assumed diagnostic tests had a 100\% detection rate to detect sampled ASFV-positive pigs, including the status Exposed ($E$), Clinical ($C$), Sub-Clinical ($S_c$), Carrier ($C_c$), Exposed(ASFV)-PRRSV(Infected) $(E_p)$, Clinical(ASFV)-PRRSV(Infected) $(C_p)$ and Detected ($D$). Pigs classified as Detected ($D$) as a result of prior internal farm surveillance were included in the evaluation of sampling efficacy; for the purpose of sampling analysis, these pigs were assumed to have the same detection characteristics as Clinical ($C$) pigs. Inclusion of Exposed ($E$) and Exposed/PRRSV co-infected ($E_p$) pigs reflected the possibility of detecting ASFV during the incubation period, since viral DNA can be detected in blood as early as $2$ days post-infection, prior to the onset of clinical signs \citep{vu_evaluation_2024, lee_pathogenicity_2021, havas_assessment_2022}. A farm was considered detected as soon as one ASFV-positive pig was identified.

\begin{figure*}
    \includegraphics[width=\linewidth]{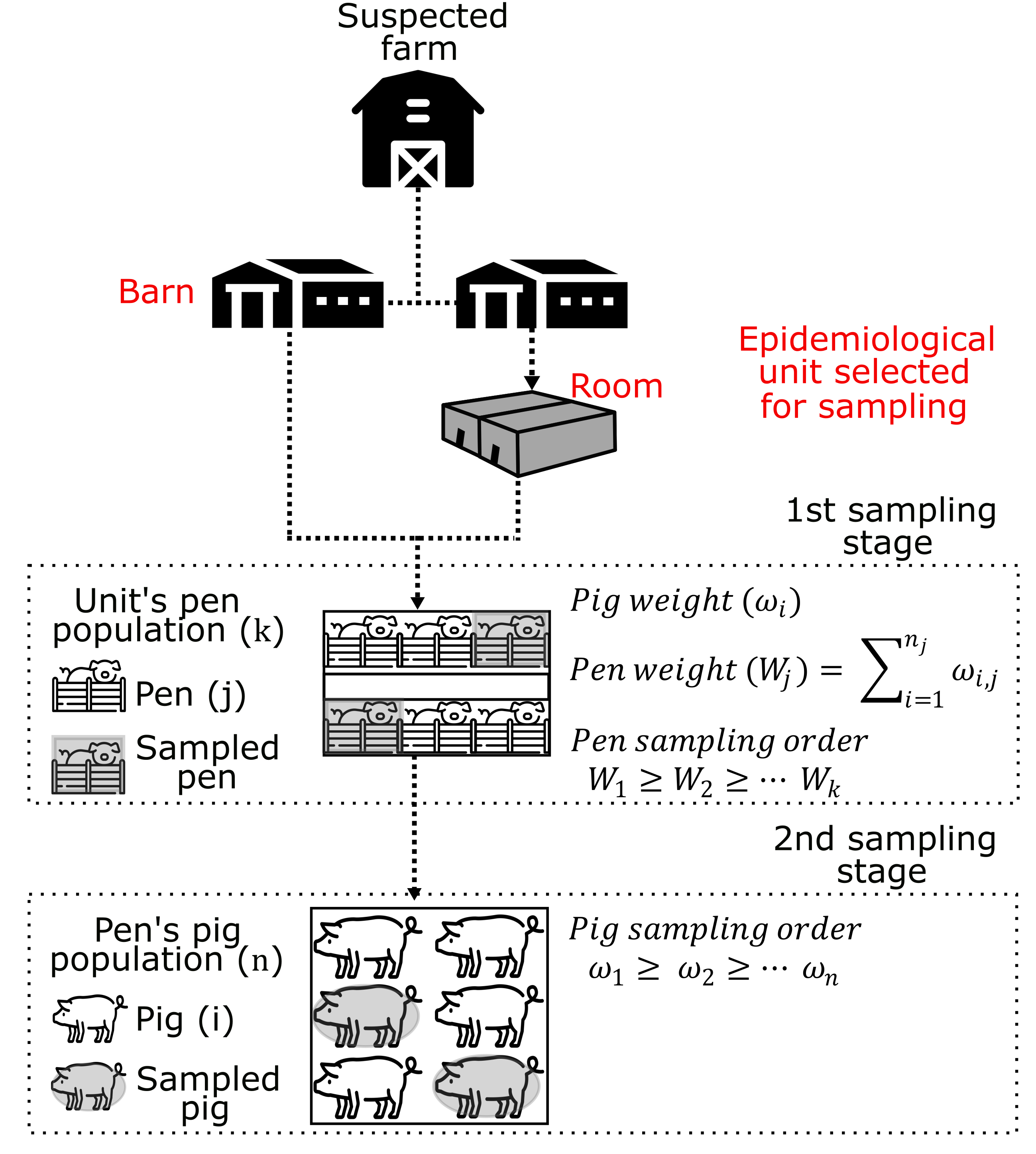}
    \caption{\textbf{ASFV sampling flow and stages.} Farm sampling can be conducted at two epidemiological units: barn or room level. Within each epidemiological unit present at the farm, a subset of pens are selected for sampling (first sampling stage), followed by a subset of pigs within each selected pen (second stage). In target pen and pig sampling process, first each pig is assigned a weight ($\omega$) drawn from a uniform distribution, reflecting the sample collector’s subjective assessment of ASFV clinical signs when co-circulating diseases are present. The total pen weight ($W$) is then calculated as the sum of $\omega$ for all pigs within pen. Finally, the selection order of pens and pigs is determined based on $\omega$ and $W$, arranged in descending order, respectively.      
    }
    \label{fig:sampling_stages}
\end{figure*}

Sampling of suspected farms was conducted at two different epidemiological unit levels: barns and rooms (Figure \ref{fig:sampling_stages}). While room-level sampling used pen population per room, barn-level sampling aggregated the pen populations across all rooms per barn. Consequently, if a fixed number of pens (e.g., five) was selected for sampling within an epidemiological unit, all selected pens were from the same room in room-level sampling, whereas for barn-level a sample of pens was taken from the available list of pens across all rooms in a barn (e.g., for a barn with room A and B, the sample would include two pens from room A and three from room B, when sample size as five) (Supplementary Material Figure S2). 

The sampling within each epidemiological unit followed a hierarchical multistage approach. In the first stage, we determined how many pens to sample from the total number of pens by epidemiological unit (Figure \ref{fig:sampling_stages}). Subsequently, in the second stage, we determined how many animals to sample from each selected pen.

\subsubsection{First sampling stage: Pen level}
We have developed four different pen sampling strategies, one random and three multistage, these include: i) targeted \& random, ii) targeted \& sequential pen selection based on a probability proportional to size (PPS) technique, and iii) targeted \& random pen selection within two clusters identified based on a PPS technique (Table \ref{tab:sampling_methods}). We assumed two possible scenarios for the pen sample size: i) the number of pens was predefined through three different scenarios (Table \ref{tab:predefined_sample_size}) and ii) the number of pens was drawn from the sample size calculation based on Cochran’s formula with a finite population correction, equation \ref{eq:sample_size}. 
\begin{equation}
n = \frac{Z^2 \cdot p \cdot (1 - p)}{E^2 \cdot \left(1 + \frac{Z^2 \cdot p \cdot (1 - p) - 1}{N \cdot E^2}\right)}
\label{eq:sample_size}
\end{equation}

\noindent where N was the estimated number of pens within an epidemiological unit (room or barn), and assuming an expected pen prevalence (p) of 1\%, a margin of error (E) of 5\%, and a confidence interval (Z) of 95\%.

\begin{table}[h!]
\centering
\caption{Sampling methods and corresponding sample size methods}
\label{tab:sampling_methods}
\resizebox{\textwidth}{!}{%
\begin{tabular}{|l|l|l|l|l|l|l|}
\hline
\multicolumn{2}{|c|}{\textbf{Sampling pen method}} & \multicolumn{2}{c|}{\textbf{Sampling animal method}} & \multirow{2}{*}{\textbf{Epidemiological unit}} & \multirow{2}{*}{\textbf{Pen sample size}} & \multirow{2}{*}{\textbf{Animal sample size}} \\ \cline{1-4}
\textbf{Primary} & \textbf{Secondary} & \textbf{Primary} & \textbf{Secondary} &  &  &  \\ \hline
 Random & - & Target    & Random & Barn or room & Pre-defined or calculated & Pre-defined or calculated \\ \hline
 Target & Random & Target & Random & Barn or room & Pre-defined or calculated & Pre-defined or calculated \\ \hline
 Target & Sequential PPS & Target & Random & Barn or room & Pre-defined or calculated & Pre-defined or calculated \\ \hline
 Target & Cluster PPS \& Random & Target & Random & Barn or room & Pre-defined or calculated & Pre-defined or calculated \\ \hline
\end{tabular}%
}
\end{table}

\begin{table}[h!]
\centering
\caption{Evaluated scenarios for the pre-defined sample size of pens and pigs}
\label{tab:predefined_sample_size}
\begin{threeparttable}
    \begin{tabular}{|l|l|l|}
        \hline
        \textbf{Scenario} & \textbf{Pens}\tnote{*} & \textbf{Animals/pen}\tnote{**} \\
        \hline
        1 & 5  & 6  \\
        \hline
        2 & 15 & 2  \\
        \hline
        3 & 30 & 1  \\
        \hline
    \end{tabular}
    \begin{tablenotes}
        \footnotesize
        \item[*] If the number of pens within an epidemiological unit was less than the number of pens required for sampling, the pen sample size was the maximum number of pens within the epidemiological unit.
        \item[**] If the number of animals within a pen was less than the number of animals required for sampling, the animal sample size was the maximum number of animals within the pen.
    \end{tablenotes}
\end{threeparttable}
\end{table}

In the random pen sampling method, pens within epidemiological units had an equal probability of being selected. In contrast, target and informative sampling methods (sequential PPS and cluster PPS) prioritized pens based on disease evidence or infection probability \citep{usda_african_2023}. Target sampling focused on pens with pigs exhibiting ASFV infection clinical signs, such as high fever, anorexia, listlessness, cyanosis, incoordination, increased pulse and respiratory rate, leukopenia, thrombocytopenia, vomiting, and diarrhea \citep{usda_african_2023}. These clinical signs were modeled as Clinical status in the ASFV transmission model. In target methods, pens were sampled sequentially until the required sample size was achieved. For example, if the sample size required was five pens and six pens contained clinical animals, only five were selected from the six. Conversely, if there were fewer than five pens with clinical animals (e.g., two pens), the remaining pens were selected randomly or based on a predefined sampling method (Table \ref{tab:sampling_methods}). Additionally, for this targeted sampling strategy, we modeled the subjective assumption that sample collectors preferentially selected pens containing true ASFV-clinical pigs when a co-circulating disease was present \citep{guinat_english_2016}. To represent this, we assumed an average error of 22\% by the sample collector during target sampling. Thus, true ASFV-positive pigs, categorized as PRRSV(Clinical)-ASFV(Clinical) and ASFV(Clinical), were assigned weights ranging from 0.2 to 0.7, drawn from a uniform distribution. Conversely, ASFV-negative but PRRSV-positive pigs, categorized as PRRSV(Clinical)-ASFV(Susceptible) and PRRSV(Clinical)-ASFV(Exposed), were assigned weights ranging from 0.2 to 0.5. Thus, the 22\% subjective sampling error was calculated as 1 - 0.35/0.45 = 0.22, where 0.45 and 0.35 represent the average weights of ASFV-positive pigs and ASFV-negative but PRRSV-positive pigs, respectively. The cumulative weights of all pigs within each pen were calculated, and pens were prioritized for sampling in descending order (Figure \ref{fig:sampling_stages}). By incorporating this weighting system, our model accounts for sampling uncertainty in the presence of a co-circulating disease, while increasing the likelihood of selecting pens with true ASFV-clinical pigs.

We complemented the target sampling strategy by integrating an informative sampling technique to provide a structured approach to sampling (Table \ref{tab:sampling_methods}). In the informative sampling strategy, pens were weighted based on their infection probability. Briefly, this infection probability represents ASFV transmission dynamics among pens, where the frequency of pen infection varies depending on the number and spatial configuration of pens, as well as ventilation characteristics (fan performance and placement) within the room. Further details are provided in \citep{safari_modeling_2024, deka2025modeling}. The weights were derived using the ASFV disease spread model, which involved seeding each pen with initial exposure $100$ times, resulting in $100 \times$ \textit{number of pens} simulations for each room across all farms. In each simulation, an Exposed pig was randomly introduced into a pen, and the infection dynamics were simulated over seven days. This approach allowed us to determine the frequency of infected animals in each pen ($p_i$), and subsequently calculate the corresponding weight ($\zeta_i$), as follows:

\begin{equation}
\zeta_i = \frac{p_i}{\sum_{i=1}^{n} p_i}
\label{eq:pen_weight}
\end{equation}

\noindent Subsequently, pens were arranged in descending order based on their $\zeta_i$ values or divided into two clusters: a high-risk cluster, containing pens where $\zeta_i$ was above the 80th percentile, and a low-risk cluster, with $\zeta_i$ below the 20th percentile. For the target \& sequential PPS method, we selected pens in descending order by $\zeta_i$ values (Table \ref{tab:sampling_methods}). In contrast, for the target \& cluster PPS method, we sampled 80\% of the pens from the high-risk cluster and 20\% from the low-risk cluster. An example of this classification was provided in Supplementary Material Figure S3.

\subsubsection{Second sampling stage: Pig level}
Animal sampling within pens were at random, except when ASFV-clinical animals were present, in which case targeted sampling was prioritized. In targeted sampling, pigs were selected based on their $\omega$ values in decreasing order (Figure \ref{fig:sampling_stages}). In addition, the animal sample size followed two strategies \ref{tab:sampling_methods}), i) a pre-defined number of sample animals per pen, and ii) calculating the animal sample size based on equation \ref{eq:sample_size}, N the estimated average of animals for each pen in the barn, and assuming p = 1\%, E = 5\%, and  Z = 95\%. 

\subsection{Outputs}

For each farm, we conducted $1000$ independent ASFV simulations for each room, randomly allocating an Exposed pig in a random pen at a random room and simulating transmission over 14 days. The model calculated sensitivity as the proportion of simulations that successfully detected at least one ASFV-positive pig out of the total simulations evaluated.

\begin{equation}
    \text{Sensitivity} = \frac{\text{Number of simulations with at least one detected pig on the farm}}{\text{Total number of simulations}}
    \label{eq:sensitivity}
\end{equation}

\noindent We evaluated the sensitivity of the sampling methods described in Table \ref{tab:sampling_methods} to identify the most effective strategy for detecting a high and moderate virulent ASFV strain while PRRSV was co-circulating within the farm. This was achieved by ranking the evaluated scenarios within each day post-infection (DPI) and sample size schema. In addition, we estimated the proportion of farms detected through passive surveillance and active surveillance triggered by pig mortality in each DPI. Finally, we examined the variation in farm detection sensitivity using three different approaches: i) assessing a moderate virulent ASFV strain, ii) evaluating scenarios with a PRRSV prevalence of 0\% and 30\%, and iii) adjusting the sampling size parameters, including an expected prevalence of 10\%, margin error of 10\% and 20\%, and confidence intervals of 90\% for both pen and animal sample sizes. All analyses were conducted in \texttt{R} \citep{r_core_team_r_2023}

\section{Results}\label{res}

\subsection{Detection by internal farm surveillance: passive and active based on mortality levels}

\begin{figure*}[!htb]
   \includegraphics[width=\linewidth]{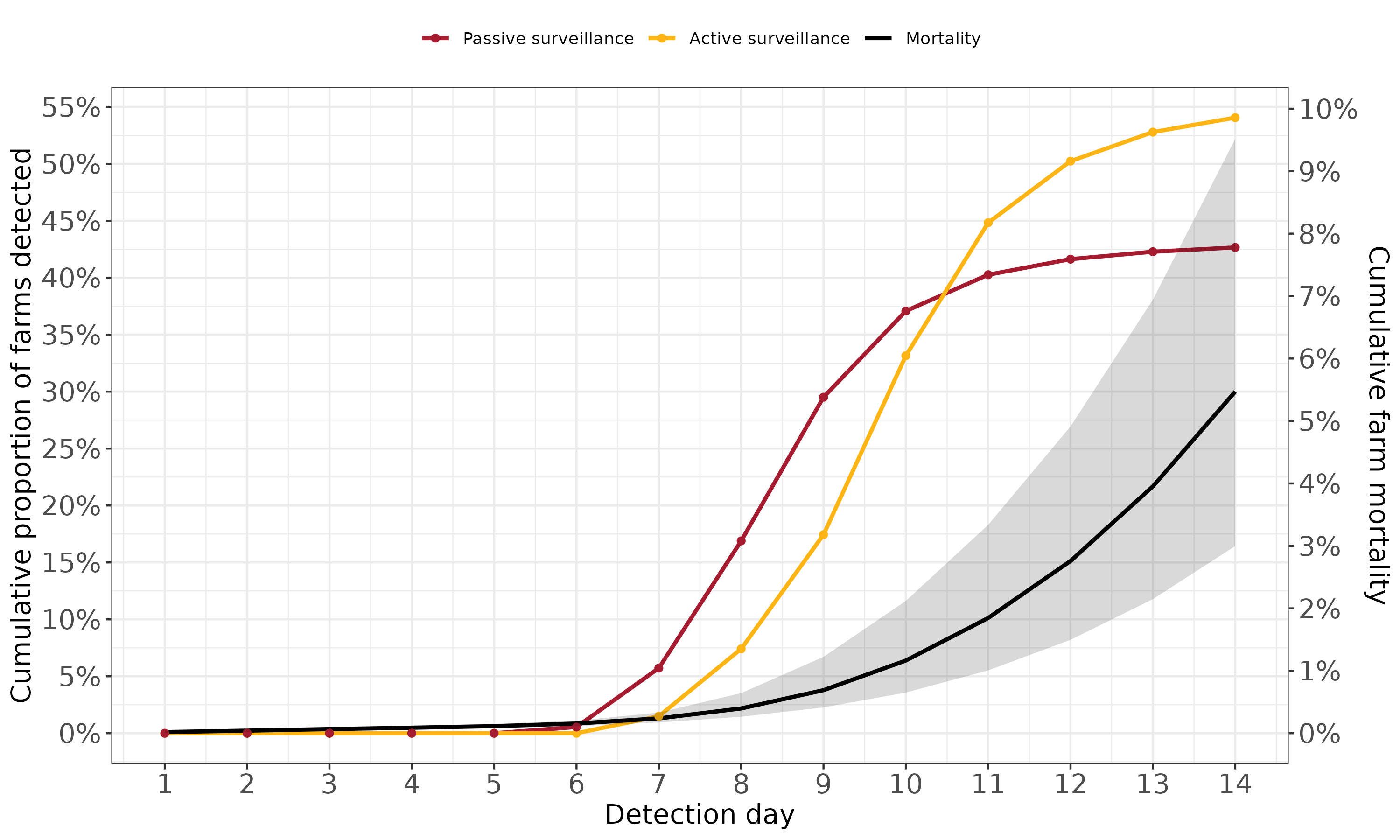}
    \caption{Cumulative proportion of infected farms with high ASFV strain detected by internal farm passive and active surveillance methods in 14 days post-infection (left y-axis), overlaid with the median cumulative mortality and its interquartile range (right y-axis). Surveillance detection curves are stratified by surveillance type, while the mortality trend is represented by a black line with a shaded ribbon indicating the 25th and 75th percentiles.}  \label{Surveillance Sensitivity}
\end{figure*}

Over the 14-day simulation period of highly pathogenic ASFV, farms were detected through either passive or mortality-triggered active surveillance at a median of 9 days (Mean: 9.6 days and IQR: 8–11)(Figure \ref{Surveillance Sensitivity}). The median proportion of simulations in which farms were detected through passive surveillance was 42.6\%, while 54.1\% were detected via active surveillance triggered by mortality levels. In 3.3\% of simulations, farms remained undetected within the 14-day period. Active surveillance identified positive farms in a median of 10 days (IQR: 9–11), while passive surveillance detected with median of 9 days (IQR: 8–10). Cumulative farm's mortality was estimated to reach a median of 0.2\% (IQR: 0.17\%–0.33\%) dead pigs by day seven and 5.45\% (IQR: 3\%–9.5\%) by day 14 (Figure \ref{Surveillance Sensitivity}).

\subsection{Detection by suspected farm surveillance}

\subsubsection{Calculated pen and animal sample size}

The simulated distribution of pens per room had a median of 32 pens (IQR: 28–40), with each pen housing a median of 34 pigs (IQR: 29–36). Of the 1,865 farms, 90\% of the barns were estimated to have a single room. Using Equation \ref{eq:sample_size} with the following parameters: Z = 95\%, E = 0.1, and p = 0.1, the median number of pens selected for sampling was 10 (IQR: 9–12) at the room level and 11 (IQR: 10–12) at the barn level. For pig sample sizes, the median was 13 pigs per pen (IQR: 11–14).

\subsubsection{Sensitivity of highly virulent ASFV detection based on sampling strategies}

\begin{figure*}[!htb]
    \includegraphics[width=\linewidth]{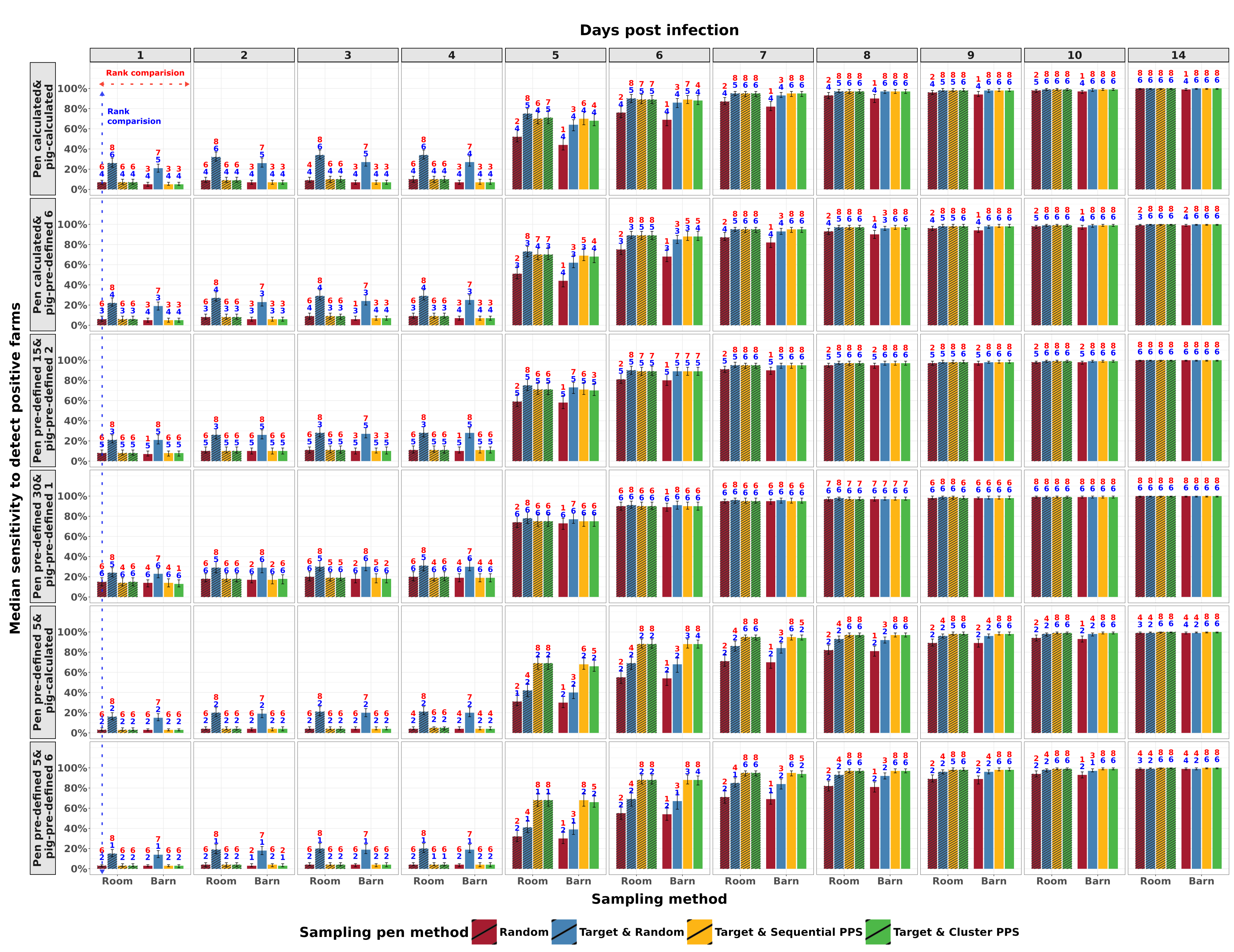}               
    \caption{Sensitivity analysis of high virulence ASFV detection through the utilization of diverse pen sampling and animal sampling methods at both barn and room levels, observed over a sequential series of days post-infection (DPI). Sampling strategies are represented by a unique color code: Random (red), target \& cluster (green), target \& random (blue), and target \& sequential (yellow). Sensitivity metrics are calculated by identifying the presence of pigs in any ASFV states Exposed, Clinical, Sub-Clinical, Carrier, PRRSV-Exposed, and PRRSV-Clinical within each epidemiological unit. Numerical rankings in descending order are depicted in red and blue to indicate the effectiveness of sampling methods. Red numbers represent the rankings (from best 8 to least effective 1) of pen sampling methods across different room and barn levels within each combined DPI and sample size schema (e.g., red horizontal line ranking comparison). Blue numbers indicate the highest-ranked methods on each specific DPI for each sampling pen method (from best 6 to least effective 1) (e.g., blue vertical line ranking comparison). Methods with identical median sensitivity values were assigned the same rank, using the upper rank value.}  
    \label{Sensitivity Room}
\end{figure*}

Our results showed that the sensitivity to detect ASFV-positive suspected farms gradually increased over time, regardless of the sampling strategy used (Figure \ref{Sensitivity Room}). At one DPI, the median sensitivity was 9\% (IQR: 4\% to 16\%) at room level and 7\% (IQR: 4\% to 15\%) at barn level. By five DPI, sensitivity increased to 68\% (IQR: 56\% to 76\%) at room level and 65\% (IQR: 51\% to 74\%) at barn level. By nine DPI, it reached 98\% (IQR: 95\% to 99\%), and by 14 DPI it peaked at 100\% (IQR: 99\% to 100\%) for both room- and barn-level sampling. Overall, sensitivity was significantly higher for room-level sampling, with a median of 2\% greater than barn-level sampling across all farms (p < 0.05) (Supplementary Material Figure S4). The difference was even more marked in farms with multiple rooms per barn, where room-level sampling achieved a median sensitivity 6\% higher than barn-level sampling (p < 0.05) (Supplementary Material Figure S5).

Examining the sensitivity across DPI, we identified three distinct performance patterns: from one to four DPI, from five to eight DPI, and after nine DPI. After nine DPI, no differences were observed for most of the sampling strategies (Figure \ref{Sensitivity Room}). Below, we describe the performance of the sampling strategies for the first two time periods.

\paragraph{Best sampling strategy} From one to four DPI, the target \& random pen sampling method consistently showed higher sensitivity, regardless of the animal or pen sample size (Figure \ref{Sensitivity Room}). During this period, its median sensitivity was 16\% (IQR: 12\% to 21\%) higher than that of all other sampling pen methods. Among the evaluated sample sizes, the predefined scheme of 30 pens with one pig per pen achieved the best overall performance, ranking as the most effective in 87\% of scenarios (Figure \ref{Sensitivity Room}) and showing a median sensitivity 10\% (IQR: 6\% to 14\%) higher than other sample size schemes. Overall, combining the target \& random pen sampling method with this predefined sample size yielded a sensitivity of 29\% (IQR: 27\% to 30\%) between one and four DPI. From five to eight DPI, this combination continued to show the best performance, with the predefined scheme of 30 pens with one pig per pen ranked as the most effective in 100\% of scenarios and showing a median sensitivity 3\% (IQR: 0\% to 11\%) higher than other sample size schemes. Although the target \& random pen sampling method performed best with this sample size scheme between five and eight DPI, its sensitivity was 0\% (IQR: 0\% to 2\%, 95th percentile: 6\%) higher than that of other pen sampling methods. In general, this combination produced a median sensitivity of 93\% (IQR: 88\% to 96\%) between five and eight DPI.

\paragraph{Second best sampling strategies} Target \& random pen sampling method combined with either a calculated pen and pig sample size scheme, a calculated pen sample size with 6 predefined pigs per pen, or 15 predefined pens with 2 pigs per pen showed the second-best performance. These sample size schemes were ranked as the best strategy in up to 12\% of scenarios from one to four DPI, and in 25\% to 28\% of scenarios from five to eight DPI. Overall, combining the target \& random pen sampling method with these sample size schemes resulted in a median sensitivity of 26\% (IQR: 24\% to 28\%) from one to four DPI and 91\% (IQR: 82\% to 95\%) from five to eight DPI.

\paragraph{Worst sampling strategies}
The lowest-performing sampling strategies were those using five predefined pens, either with a calculated pig sample size or with six predefined pigs per pen. These strategies ranked last in 100\% of scenarios from one to four DPI and ranked best in only 21\% of scenarios from five to eight DPI (Figure \ref{Sensitivity Room}). Within this sample size scheme, the target \& random pen sampling method performed best from one to four DPI, with a median sensitivity of 19\% (IQR: 17\% to 20\%). From five to eight DPI, the target \& sequential PPS and target \& cluster PPS methods consistently outperformed the others, achieving a median sensitivity of 91\% (IQR: 83\% to 95\%), representing a 15\% increase over the target \& random pen sampling method during this period.

\subsubsection{Pig sample size based on sampling strategies}

\begin{figure*}[!htb]
   \includegraphics[width=\linewidth]{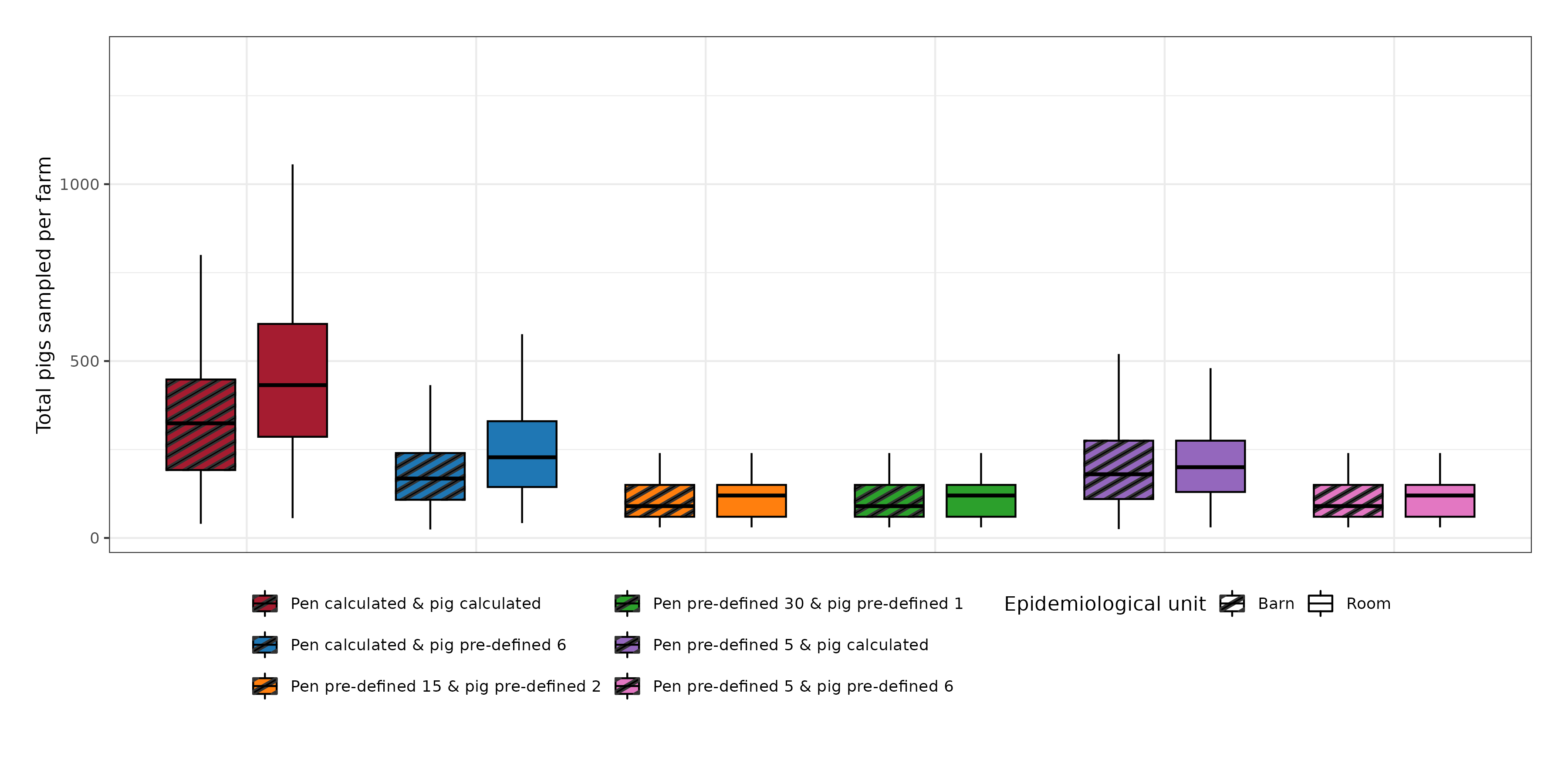}
    \caption{Number of pigs sampled per farm across different sample size approaches. The Y-axis represents the median number of pigs sampled per farm, while the X-axis denotes the epidemiological unit used for sampling (e.g., barn or room). Different colors highlight the different pen and pig sampling method.}  
    \label{Pigs_sample_size}
\end{figure*}

The total number of pigs sampled per farm varied based on the selected sample size and the epidemiological unit used (i.e., barn or room). Overall, room-level required in median  0\% (IQR: 0\% to 3.5\%) more pigs than barn-level sampling regardless of the sampling strategy (Figure \ref{Pigs_sample_size}). However, this difference was more pronounced among farms with barns with more than one room, where the median number of pigs sampled increased by 100\% (IQR: 69\% to 100\%). Across all evaluated scenarios, the strategy that predefined sampling to 30 pigs per epidemiological unit yielded the lowest total number of pigs sampled per farm, with a median of 120 pigs (IQR: 60 to 120) per sampling event.

Using the strategy with 30 pigs predefined per epidemiological unit as a reference, going from a scenario where both pen and pig sample sizes were calculated increased the number of pigs sampled by 303\% (IQR: 294\% to 340\%) at the room-sampling level and 202\% (IQR: 193\% to 261\%) at the barn-sampling level (Figure \ref{Pigs_sample_size}). When the pen sample size was calculated and the number of pigs per pen was fixed at six, the increase was 120\% (IQR: 107\%–140\%) at the room level and 80\% (IQR: 60\%–100\%) at the barn level. Similarly, predefining five pens and calculating the number of pigs per pen resulted in a 83\% increase (IQR: 67\%–100\%) at both the room and barn levels.

\begin{figure*}[!htb]
   \includegraphics[width=\linewidth]{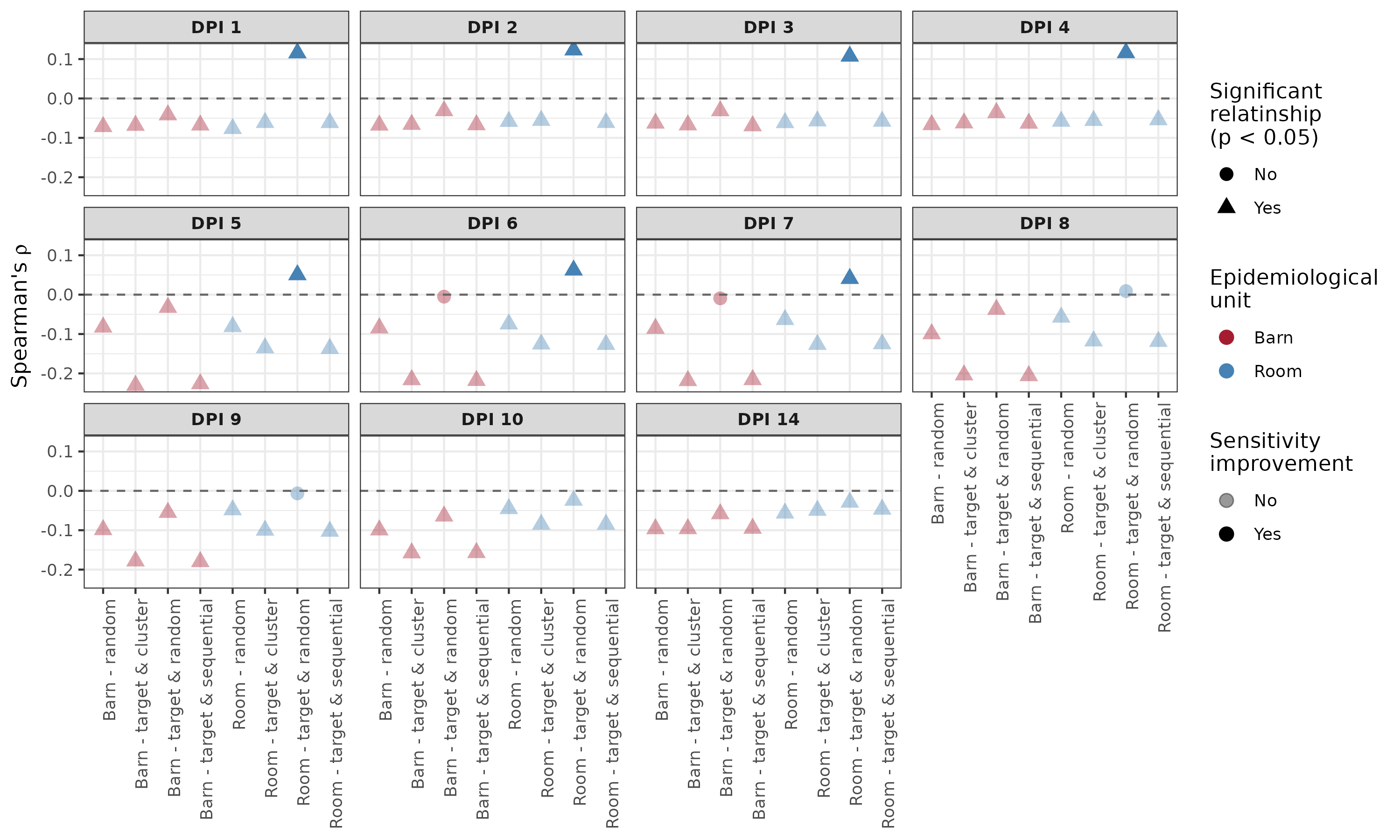}
    \caption{Spearman correlation between sensitivity and the total number of pigs sampled, stratified by epidemiological unit (barn or room) and pen sampling method, across days post-infection (DPI). Each point represents a unique combination of sampling strategy and epidemiological unit, with panel showing individual DPI. Shapes indicate whether the correlation was statistically significant ($p <$ < 0.05). Horizontal dashed lines represent no correlation ($\rho$ = 0).}  \label{Pigs Sensitivity Days}
\end{figure*}

When analyzing the correlation between sensitivity and the total number of pigs sampled (Figure \ref{Pigs Sensitivity Days}), we observed that during the first seven DPI, the target \& random method performed better with larger sample sizes. However, a statistically significant association was only detected at the room level on days one to seven ($p <$ 0.05). None of the other sampling methods or DPI showed a significant improvement in sensitivity with increasing sample size across the evaluated days and epidemiological units ($p >$ 0.05).

\paragraph{Sensitivity variation by sample size calculated parameters} Calculated sample size scenarios with an expected prevalence of 10\% resulted in higher median sensitivity values than the predefined scenario of sampling 30 pens with one pig per pen during the first five DPI, and showed similar values thereafter (Supplementary Material Figure S6). During these first five days, the calculated scenarios achieved a median sensitivity that was 21\% higher (IQR: 11\% to 45\%) than the 30 pen predefined sampling strategy. However, this increased sensitivity came at the cost of sampling substantially more animals, with a median ranging from 800 to 2,094 pigs, representing seven to 17 times more animals than the median of 120 pigs sampled in the 30 pen predefined scenario.

\subsubsection{Sensitivity variation influenced by PRRSV prevalence}

ASFV detection sensitivity in response to PRRSV prevalence was directly influenced by whether exposed pigs were detected. This effect was driven by a higher probability of selecting PRRSV-clinical pigs that were also recently exposed to ASFV ($E_p$). Because these pigs exhibited PRRSV clinical signs, they were more likely to be targeted, inadvertently increasing the likelihood of detecting ASFV. To evaluate this effect, we explored two distinct scenarios: one in which exposed pigs were detectable during sampling, and another in which they were not. When exposed pigs were detectable, the overall difference in sensitivity due to increasing PRRSV prevalence was 2\% (IQR: -1\% to 9\%). In contrast, when exposed pigs were not detectable during sampling, the difference was 0\% (IQR: -1\% to 0\%), regardless of the DPI or the sampling strategy used (Supplementary Material Figure S7).

\begin{figure*}[!htb]
   \includegraphics[width=\linewidth]{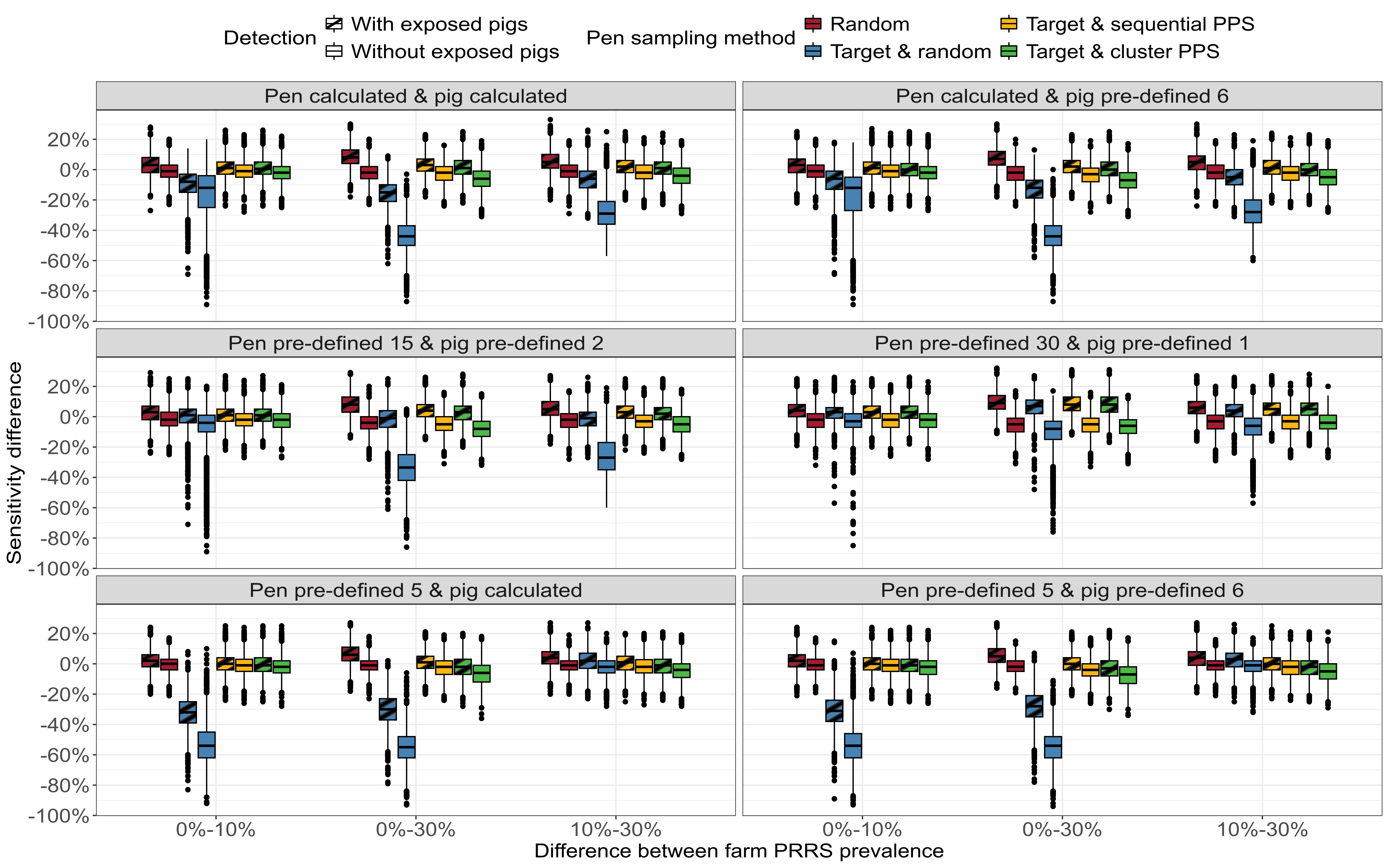}
    \caption{Difference in sensitivity between sampling strategies with varying PRRSV prevalence levels at five DPI. Each boxplot shows the distribution of sensitivity differences between two farm-level prevalence scenarios (0\%–10\%, 0\%–30\%, and 10\%–30\%). Panels represent combinations of pen and pig-level sample size methods. Colors indicate the pen sampling method used. Positive values indicate higher sensitivity under the higher prevalence condition.}  \label{prrs_prev_diff}
\end{figure*}

However, we observed variations in interactions between sampling strategies at specific DPI. To facilitate the comparison, five DPI was selected because it showed higher sensitivity than earlier time points and more pronounced differences between sampling strategies (Figure~\ref{Sensitivity Room}). When using a small number of pens, such as sampling five pre-defined pens, increasing PRRSV prevalence had a minimal effect on the sensitivity of the random, sequential PPS, and cluster PPS methods (Figure~\ref{prrs_prev_diff}). For these methods, the median sensitivity differences was 1\% (IQR: -4\% to 5\%) with detectable Exposed pigs, and -2\% (IQR: -7\% to 2\%) without detectable Exposed pigs. In contrast, sensitivity was drastically reduced when using the target \& random pen sampling method. When exposed pigs were detectable, the sensitivity differences was -31\% (IQR: -38\% to -24\%) as PRRSV prevalence increased from 0\% to 10\%, and -29\% (IQR: -36\% to -22\%) when increasing from 0\% to 30\%. When exposed pigs were not detectable, the reduction was more severe. The sensitivity differences was -54\% (IQR: -62\% to -45\%) as PRRSV prevalence increased from 0\% to 10\%, and -55\% (IQR: -62\% to -48\%) when increasing from 0\% to 30\%. Thus, sampling a small number of pens may lead to a median loss in sensitivity between 29\% and 55\% when using the target \& random pen sampling method, particularly if a misleading co-circulating disease such as PRRSV reaches a prevalence of 30\%.

We observed different results when increasing the number of pens used for sampling, independent of the pen sampling method (Figure~\ref{prrs_prev_diff}). For example, in the scenario using 30 pens, when exposed pigs were detectable, the sensitivity difference was 3\% (IQR: -1\% to 6\%) as PRRSV prevalence increased from 0\% to 10\%, and 7\% (IQR: 2\% to 11\%) when increasing from 0\% to 30\%. When exposed pigs were not detectable, the sensitivity difference was -3\% (IQR: -7\% to 2\%) as PRRSV prevalence increased from 0\% to 10\%, and -8\% (IQR: -15\% to -3\%) when increasing from 0\% to 30\%. Thus, at a co-circulating disease prevalence of 30\%, sampling a higher number of pens may increase sensitivity by a median of 7\% when exposed pigs are detectable, or result in a 8\% median reduction in sensitivity when they are not. This reduction was approximately seven times smaller than the sensitivity loss observed when sampling only five pens.

\subsubsection{Sensitivity variation by a moderate virulent ASFV strain}

\begin{figure*}[!htb]
   \includegraphics[width=\linewidth]{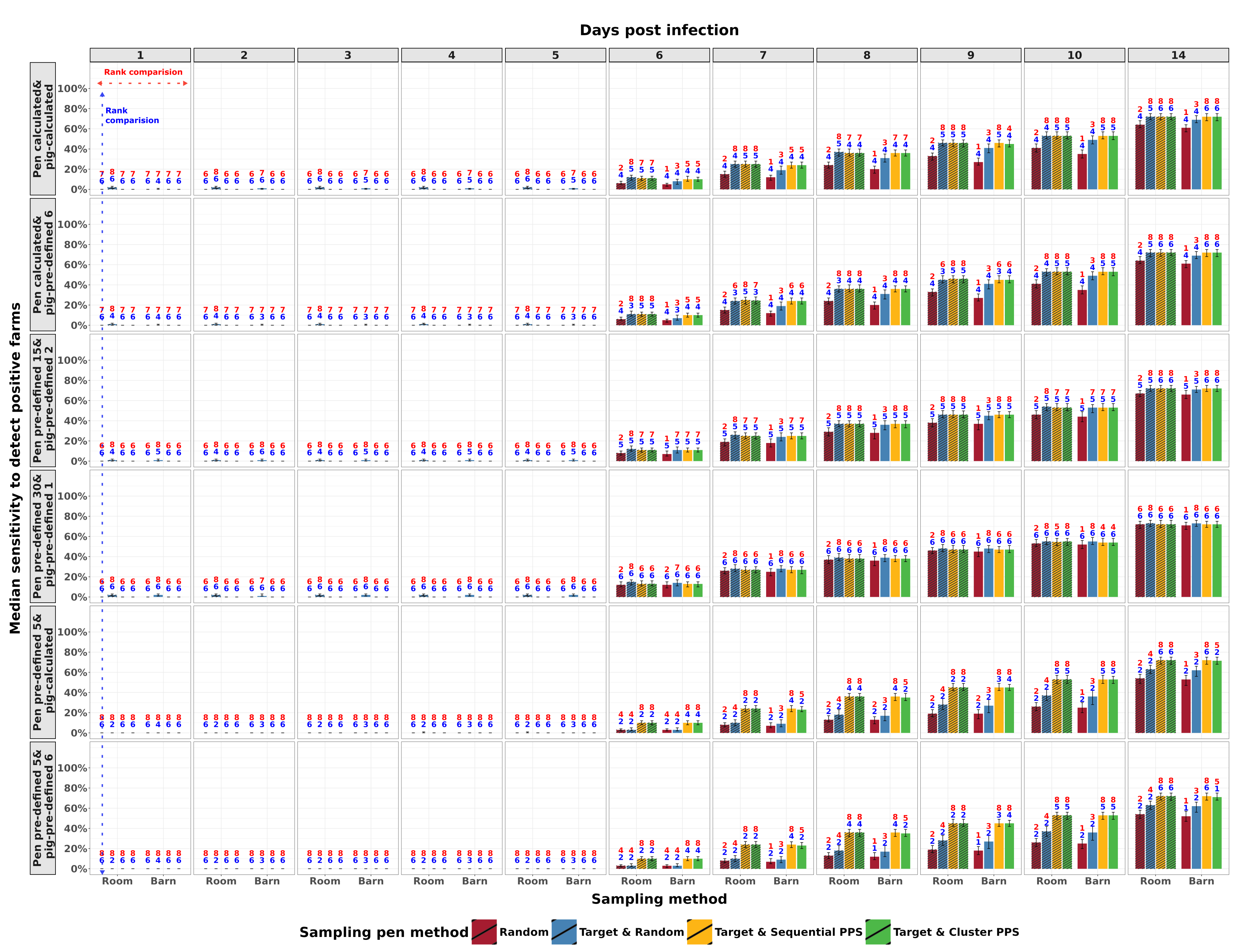}
    \caption{Sensitivity analysis of ASFV detection through the utilization of diverse pen sampling and animal sampling methods at both barn and room levels, observed over a sequential series of days post-infection (DPI) for a moderate virulent ASFV strain. Sampling strategies are represented by a unique color code: Random (red), target \& cluster (green), target \& random (blue), and target \& sequential (yellow). Sensitivity metrics are calculated by identifying the presence of any of the states Exposed, Clinical, Sub-Clinical, Carrier, PRRSV-Exposed, and PRRSV-Clinical within each epidemiological unit. Additionally, numerical rankings in descending order are depicted in red and blue to indicate the effectiveness of sampling methods. Red numbers represent the rankings (from best 8 to least effective 1) of pen sampling methods across different room and barn levels within each combined DPI and sample size schema (e.g., red horizontal line ranking comparison). Blue numbers indicate the highest-ranked methods on each specific DPI for each sampling pen method (from best 6 to least effective 1) (e.g., blue vertical line ranking comparison). Methods with identical median sensitivity values were assigned the same rank, using the upper rank value.}  
  \label{Pigs Sensitivity Low Sttrain}
\end{figure*}

At 14 DPI, only 36\% of simulations identified farms infected with the moderate-virulence ASFV strain. Extending the simulation period to 60 DPI increased the detection rate to 89\% (Supplementary Material Figure S8). Within this 60-day window, the median detection time was 16 days (IQR: 13 to 20). Similar to high-virulent ASFV, the number of suspected farms infected with moderate-virulent ASFV gradually increased over time. However, the detection sensitivity at the farm level was low, regardless of the sampling strategy used (Figure \ref{Pigs Sensitivity Low Sttrain}). From one to five DPI, the median sensitivity was 0\% (IQR: 0\% to 0\%) at both the room and barn levels. By six DPI, sensitivity increased to 10\% (IQR: 6\% to 13\%) at the room level and 9\% (IQR: 5\% to 12\%) at the barn level. At ten DPI, it rose to 50\% (IQR: 44\% to 56\%) and 50\% (IQR: 41\% to 55\%) at the room and barn levels, respectively. By 14 DPI, sensitivity reached 69\% (IQR: 65\% to 73\%) at both levels. Similar to high-virulent, there was small, but significative difference in the sensitivity levels between room and barn (Supplementary Material Figure S9).

The performance trends of the sampling strategies for the moderate-virulent ASFV strain was similar to that observed for the high-virulent strain (Figure \ref{Pigs Sensitivity Low Sttrain}). The sampling approach using 30 predefined pens with one pig per pen was the most effective, ranking best in 100\% of the evaluated scenarios. In contrast, sampling only five predefined pens, either by selecting six fixed pigs or by calculating the pig sample size, consistently showed the worst performance, ranking lowest in 100\% of the scenarios across all DPI. When a high number of pens was used, such as with calculated pen sample sizes or when 15 to 30 pens were predefined, the target \& random pen sampling method consistently outperformed all other approaches for detecting moderate-virulent ASFV infections. However, the advantage was moderate, with a median sensitivity only 1\% (IQR: 0\% to 3\%) higher than other methods. Larger differences among pen sampling methods were observed when using a low number of pens, such as five pens. In these cases, the sequential PPS and cluster PPS pen sampling methods showed better performance, with a median sensitivity 4\% (IQR: 0\% to 13\%) higher from one to 14 DPI and 12\% (IQR: 7\% to 19\%) higher from five to 14 DPI, compared to the target \& random method.

\section{Discussion}

Our study evaluated the effectiveness of 48 room- and barn-level sampling strategies for detecting high- and moderate-virulence ASFV in growing pig population in the U.S. Combined passive and mortality-based surveillance detected the high-virulence strain in 97\% of cases within 14 DPI, but only 36\% for the moderate-virulence strain. Whilst sampling pigs in suspected farms, we found that, regardless of the sampling schema, detection sensitivity increased over time. The target \& random pen sampling method, when combined with a sample size of 30 pens and one pig per pen, was the most effective for both high- and moderate-virulence ASFV. This approach required relatively few animals (120 pigs/farm) and showed minimal loss of sensitivity when PRRSV prevalence increased, which can mislead targeting of truly ASFV-clinical pigs (1\% to 15\% sensitivity reduction). In contrast, sampling only five pens with six or more pigs per pen consistently yielded the lowest performance for detecting infected farms with either ASFV strain. Under this five-pen scenario, the target \& random method underperformed compared with informative pen sampling methods (sequential PPS and cluster PPS) after four DPI, and its sensitivity was markedly reduced (29\% to 55\% sensitivity reduction) when PRRSV prevalence increased. These findings underscore the challenges of detecting ASFV-positive farms in the early stages of infection and provide critical guidance for selecting strategies that maximize detection probability in growing pig farms.

Our results showed that passive or mortality mediate surveillance was able to detect high virulent ASFV strain within 9 (IQR: 8 to 11) days, aligning with South Korea reports, where infected fattening farms were detected on average 9.5 days (95\% CI: 7.9 to 11.3) \citep{yoon_estimating_2023}. Although not evaluated in this study, variability in farm-level surveillance efficacy may delay detection times, as ASFV detection has been reported to take several weeks \citep{mur_understanding_2018, lamberga_african_2020}. Strengthening internal farm surveillance and implementing appropriate sampling protocols are therefore critical for ensuring timely detection, triggering rapid control actions, and reducing the risk of transmission from undetected farms.

To maximize farm-level ASFV detection sensitivity while minimizing the number of pigs sampled, the most effective strategy was to test one pig from each of 30 pens (or from all pens if fewer were available). When combined with targeted sampling of pens showing clinical signs and random selection in their absence, this approach achieved a median sensitivity of 29\% within four DPI and 93\% thereafter. Its superior performance is explained by within-barn transmission dynamics, ASFV spreads faster within pens than between pens, meaning pigs housed with an infected animal are more likely to be exposed sooner than those in neighboring pens. Therefore, when only a few pens are infected and clinical signs are absent, increasing the number of pens sampled maximizes the probability of detecting exposed pigs \citep{murato_evaluation_2020}. Another strength of this strategy is its practicality. Unlike more complex approaches such as statistically calculated sample sizes or informative pen sampling methods (sequential PPS and cluster PPS) that require prior modeling, the target \& random method with 30 pens and one pig per pen is simple, field-ready, and easy to implement. Its straightforward design eliminates the need for additional data processing or specialized tools, enabling timely decision-making.

A previous study evaluating sampling strategies for classical swine fever (CSF) demonstrated that selecting pens located in the corners and center of the barn, and sampling five pens, could be more efficient than a random approach \citep{murato_evaluation_2020}. Similarly, we demonstrated that sampling five pens was effective in detecting high- and moderate-virulent ASFV strains when used in conjunction with informative pen selection methods (sequential PPS or cluster PPS). In these informative methods, the pens were selected from strategic areas such as the corners of the barn  (Supplementary Material Figure S3), based on the simulated within-barn spread dynamic \citep{deka2025modeling, safari_modeling_2024}. However, sampling only five pens reduced detection performance compared with sampling 30 pens. Similarly to the CSF study \citep{murato_evaluation_2020}, our results also indicate that reducing the number of sampled pens, even when the total number of pigs sampled within the barn remains the same, decreases detection performance. Nevertheless, sampling fewer pens (e.g., five) may be necessary when logistical constraints limit sampling capacity \citep{sykes_estimating_2023, galvis2025estimating}. In this context, conducting multiple sampling events spaced a few days apart can help compensate low sensitivity in detection early ASFV infection. Because the probability of detection increases markedly after five DPI, a second sampling event five days later is likely to detect infected farms, raising the median sensitivity around to at least 68\%.

Overall, the difference between barn- and room-level sampling was minimal (75th percentile: 3.5\%), primarily because of the distribution of farms with multiple rooms per barn. However, in farms with multiple rooms per barn, room-level sampling may be beneficial, increasing median sensitivity by 6\%, although it also doubles the number of animals sampled. When adjusting sample size parameters, increasing the number of samples did not substantially improve sensitivity compared with the scenario of sampling 30 pens with one pig per pen. A sevenfold increase in sample size was required to achieve a significant improvement in sensitivity, which is likely a challenge to be implemented. Therefore, the scenario of sampling 30 pens with one pig per pen, which requires a median of 120 pigs per farm, provides an adequate sample size while maintaining high sensitivity.

A noteworthy finding was the effect of PRRSV co-circulating on the detection of ASFV-infected pigs. When five pens were sampled and the prevalence of PRRSV was high (e.g., 30\%), the performance of the target \& random pen sampling method was negatively impacted. The reduction was due to increased error in selecting pens that contained ASFV-infected pigs, resulting in a loss of detection sensitivity ranging from 29\% to 55\%. In contrast, even with the co-circulation of PRRSV, increasing the number of pens sampled (e.g., 30 pens) improved the detection sensitivity across all pen-level sampling methods, driven by a higher probability of selecting truly infected pens. In our scenarios, we allowed for the detection of PRRSV-positive clinical pigs that had recently been exposed to ASFV. These pigs were more likely to be targeted during sampling due to clinical signs associated with PRRSV. However, if ASFV exposed pigs could not be detected during sampling, increasing PRRSV prevalence reduced detection sensitivity across all sampling strategies (Figure ~\ref{prrs_prev_diff} and Supplementary Material Figure S7). This is because a higher PRRSV prevalence increases the likelihood of sampling pigs that are not ASFV-infected, thereby producing false positives within pens. In the U.S., the introduction of ASFV will ultimately occur in a population exposed to endemic pathogens; thus, sampling strategies should account for the detection challenge that co-circulation will pose. Our results show that increasing the number of sampled pens and expanding training programs, such as the U.S. Certified Swine Sample Collector (CSSC) initiative \citep{secure_pork_supply_plan_certified_2024}, would enhance the chances of identifying ASFV-positive animals during early surveillance.

A moderate-virulence ASFV strain presented in a subclinical form is more difficult to recognize in the field \citep{gallardo_african_2015}, likely leading to longer detection delays. We emulated this detection delay by assuming a 50\% reduction in the probability of passive farm detection compared to a highly virulent ASFV strain. We found that only 36\% of simulations identified a positive farm within 14 DPI through internal farm surveillance. When the observation window was extended to 60 days, the median detection time was 16 DPI; however, in some cases, detection took several weeks to months (Supplementary Material Figure S8). This detection timeline aligns with the reported onset of clinical signs, typically occurring between 10 and 20 DPI, for moderate-virulence ASFV strains \citep{sanchez-vizcaino_update_2015}. Additionally, we observed cumulative mortality reaching a median of 30\% within 60 days (Supplementary Material Figure S8), consistent with previously reported mortality rates ranging from 30\% to 70\% after 20 DPI for moderate virulent ASFV \citep{gallardo_african_2015}. For the sampling strategies in the suspected farm surveillance, we estimated detection rates still below 100\% at 14 DPI. The performance of the sampling strategies, based on the sample size and pen selection method, followed a pattern similar to that observed for the highly virulent ASFV strain. When using the best-performing sampling strategy, which involves sampling 30 pens with one pig per pen utilizing a combination of targeted and random pen selection, it identified infected farms with moderate-virulence ASFV strain in a median of five days earlier than internal farm surveillance. These results highlight the importance of suspected farm surveillance for earlier detection of moderate-virulence ASFV infections. Despite this advantage, overall detection sensitivity remained below 50\% during the first 10 DPI, suggesting that multiple sampling rounds are likely necessary to identify infected farms. If only two sampling rounds are feasible, we showed that conducting the second round around 10 DPI would improve the probability of detection. This timing was selected because, in cases where the first sampling yields negative results, likely due to being conducted during the early stage of infection when within-farm prevalence is still low, a second sampling after approximately 10 days would provide a higher likelihood of detecting ASFV-positive animals.

\section{Limitations ad further remarks}
This study has limitations, and its results should therefore be interpreted with caution. We only considered growing pig farms, which typically have uniform barn designs, pen configurations and population dynamics (all-in/all-out) compared to breeding pig farms. Additional analyses are needed to adapt and validate our methodology for breeding farms and boar stud farms, where barn layout and operational differences may yield results that differ from those observed in growing farms. Our pen distribution by barn was based on pen metrics reported in the literature, assuming an even distribution of the reported animal capacity across pens. However, this assumption may not accurately reflect all farms, as pen layouts and animal distributions vary. This simplification may influence the modeled ASFV spread and the performance of the evaluated sampling strategies. 

We modeled internal farm passive and active surveillance under the assumption that stakeholders are aware of ASFV circulation, making testing and notification mandatory \citep{usda_african_2023}. While the U.S. swine industry is undergoing preparedness training for a potential ASFV outbreak \citep{secure_pork_supply_plan_certified_2024}, the actual effectiveness and responsiveness of the internal farm surveillance remain unknown. Although conditions in European and African countries differ from those in the U.S., ASFV detection has proven to be a challenge due to unreported suspected cases for various reasons \citep{penrith2024african, vergne_attitudes_2016, costard_small-scale_2015, guinat_english_2016}. In addition, we assumed that 2\% animal mortality within two weeks would trigger internal farm active surveillance. However, such mortality levels may also be caused by other endemic diseases, such as PRRSV, which can reach similarly high mortality rates \citep{osemeke_economic_2025}. As a result, the detection timelines estimated in our study likely reflect an optimistic scenario. To produce more robust and realistic estimates, additional information on stakeholder behavior and the operational performance of official surveillance systems is needed. In the sampling design, sample sizes were estimated using Cochran’s. Although suitable for prevalence estimation, alternative methods such as hypergeometric or probabilistic approaches may provide an alternative for demonstrating disease freedom and could be considered in future model developments \citep{cameron_new_1998}.

Although PRRSV was included as a co-circulating disease, we did not model its infection dynamics or assess how co-infection might influence ASFV transmission or internal farm surveillance. For instance, by increasing host susceptibility and accelerating ASFV spread, or accelerating pig mortality levels and detection \citep{rajkhowa_natural_2024, cabezon_african_2017}. A comprehensive approach that explicitly incorporates the transmission dynamics of both pathogens could provide insights and improve the relevance of the findings for ASFV surveillance strategies. Additionally, while we explored different PRRSV prevalence levels (e.g., 0\%, 10\%, and 30\%) at the farm level, it remains uncertain whether these values accurately reflect real conditions in the U.S., given the limited studies estimating farm-level prevalence in growing pig population \citep{galvis_modelling_2022}.

In our study we considered the detection of ASFV-exposed pigs. However, in the real-world scenario, ASFV-exposed pigs may go undetected if sampling occurs shortly after infection, particularly within the first two days post-exposure \citep{vu_evaluation_2024, lee_pathogenicity_2021, havas_assessment_2022}. This variability in the probability of detecting ASFV-exposed pigs during early stages represents a limitation of the present study. Future modeling efforts may benefit from incorporating an "incubation" pig state to better distinguish recently exposed pigs and improve the representation of detection dynamics. Despite these limitations, our study highlights the differences in detecting ASFV-exposed pigs and offers a better understanding of the impact of co-infections on sampling performance. 

Regarding the weight assigned to ASFV- and PRRSV-clinical pigs during targeted sampling, we assumed that sample collectors would have on average a 22\% reduction in selection likelihood to identify pigs truly infected with ASFV. However, this weighting was based on subjective judgment, and our assumptions may have been either overly optimistic or overly conservative \citep{guinat_english_2016}. To generate robust conclusions about targeted sampling in the presence of co-circulating diseases, a better understanding of sample collector behavior and potential misclassifications during sampling events is needed.

We found that the target \& random pen sampling method was the most effective and straightforward approach for detecting ASFV. However, informative sampling methods such as sequential PPS and cluster PPS also performed well. It is possible that other pen sampling strategies not evaluated in this study could offer even better performance. For example, targeting and increasing sampling intensity in barns or pens where disease introduction is more likely, as suggested in previous studies \citep{lamberga_african_2020}. While more complex sampling designs may theoretically improve detection, their practical implementation must also be considered. Increased complexity can create logistical challenges and may reduce overall sampling effectiveness by confusing sample collectors during field application.

We demonstrated that sampling 30 pigs, distributed across 30 pens, was the most effective configuration. However, it is possible that a smaller sample size, while maintaining the exact configuration of one pig per pen, could yield similar results. Reducing the number of pens sampled may help decrease logistical burdens by saving time and resources required for sample collection. Further analysis focused on this aspect could support the development of more cost-effective ASFV surveillance strategies.

Despite the limitations of this study, our findings provide practical, evidence-based insights from 1,865 farm data that strengthen ASFV surveillance frameworks in the U.S. and other ASFV-free countries. In particular, the demonstrated effectiveness of a simple pen-level sampling strategy, which involves sampling one pig from each of up to 30 pens, aligns well with the Red Book’s emphasis on feasible and rapid field implementation \citep{usda_african_2023}. While informative pen sampling methods may perform better under certain conditions, our results highlight the consistent performance and operational simplicity of the 30-pen approach. The findings also underscore the importance of repeated sampling in addressing detection challenges during the early DPI. Incorporating these results into operational surveillance guidelines could enhance the sensitivity of early detection and reduce the risk of undetected spread. As USDA and state animal health officials continue to refine ASFV preparedness plans, the integration of modeling-based sampling recommendations can support more adaptive, efficient, and cost-effective surveillance during both preparedness and emergency response phases.

\section{Conclusion}
This study evaluated a wide range of room- and barn-level sampling strategies for early ASFV detection in commercial growing pig farms in the U.S., highlighting the trade-offs between sensitivity and the feasibility of different sample size. Sampling 30 pens (or all pens if fewer were available) with one pig per pen, combined with the target \& random pen selection method, consistently achieved the highest detection sensitivity while minimizing the number of pigs tested, providing a practical and scalable option for field implementation. This approach was particularly effective during the early days post-infection and maintained strong performance under both high- and moderate-virulence ASFV scenarios, even in the presence of co-circulating diseases such as PRRSV. In contrast, strategies that sampled only five pens showed lower detection sensitivity, and when combined with the target \& random method, performed worse than those using informative pen sampling methods. Overall, our findings emphasize the value of increasing the number of pens sampled and optimizing pen selection methods when low-pen sampling is unavoidable, to improve detection performance under real-world constraints. Incorporating these sample size and pen selection strategies into surveillance protocols can strengthen early detection capacity and improve preparedness for ASFV incursions in the U.S., as well as for other foreign animal diseases with similar transmission dynamics.

\section*{Funding}
This work was also supported by the Foundation for Food \& Agriculture Research (FFAR) award number FF-NIA21-0000000064 and from the National Institute of Food and Agriculture, Data Science for Food and Agricultural Systems (DSFAS), A Novel Multilevel Model of Swine Disease Spread to Assess the Effectiveness and Feasibility of African Swine Fever Control and Eradication Strategies, Grant/Award Numbers:2024-67021-43841.

\bibliographystyle{elsarticle-harv}
\bibliography{references1}
\end{document}